\documentclass[journal]{IEEEtran}
\usepackage{graphicx}
\usepackage{pifont}
\usepackage{makecell,rotating}
\usepackage[ruled,vlined]{algorithm2e}
\usepackage{algpseudocode}
\usepackage{upgreek}
\usepackage{color}
 
\usepackage{amsmath,amsthm,amssymb,amsfonts}
\usepackage{epstopdf}
\usepackage{bm}
\usepackage{eucal}
\usepackage{verbatim}
\usepackage{subfigure}
\usepackage{xspace}
\usepackage{setspace}
\usepackage{calrsfs}  
\usepackage{multirow}
\usepackage{balance}
\usepackage{booktabs,color}
\usepackage{tabularx}
\usepackage{xcolor}
\usepackage{bbding,enumitem}
\usepackage{caption}
\def\BibTeX{{\rm B\kern-.05em{\sc i\kern-.025em b}\kern-.08em
    T\kern-.1667em\lower.7ex\hbox{E}\kern-.125emX}}

\newcommand{\rf}{\textsf{RF}\xspace}
\newcommand{\varcnn}{\textsf{Var-CNN}\xspace}
\newcommand{\df}{\textsf{DF}\xspace}
\newcommand{\netclr}{\textsf{NetCLR}\xspace}
\newcommand{\tiktok}{\textsf{Tik-Tok}\xspace}

\newcommand{\tmwf}{\textsf{TMWF}\xspace}
\newcommand{\bapm}{\textsf{BAPM}\xspace}
\newcommand{\ours}{\textsf{ARES}\xspace}
\newcommand{\transformer}{\textsf{Trans-WF}\xspace}
\newcommand{\eg}{\emph{e.g.,}\xspace}
\newcommand{\ie}{\emph{i.e.,}\xspace}
\newcommand{\first}{\textsf{(i)}\xspace}
\newcommand{\second}{\textsf{(ii)}\xspace}
\newcommand{\third}{\textsf{(iii)}\xspace}

\newcommand\xinhao[1]{{\color{black} #1}}
\newcommand\update[1]{{\color{black} #1}}
\definecolor{mycolor1}{HTML}{d7191c}
\definecolor{mycolor2}{HTML}{2b83ba}
\definecolor{mycolor3}{HTML}{14884e}

\newcommand{\xmark}{\textcolor{red}{\ding{55}}} 
\newcommand{\halfcmark}{\textcolor{mycolor2}{\ding{52}\rotatebox[origin=c]{-9.2}{\kern-0.7em\ding{55}}}} 
\newcommand{\cmark}{\textcolor{mycolor3}{\ding{52}}}     

\author{Xinhao Deng,~\IEEEmembership{Student Member,~IEEE,}, Xiyuan Zhao, Qilei Yin, Zhuotao Liu, Qi Li,~\IEEEmembership{Senior~Member,~IEEE,} Mingwei Xu,~\IEEEmembership{Senior~Member,~IEEE,} Ke Xu,~\IEEEmembership{Fellow,~IEEE,} Jianping Wu,~\IEEEmembership{Fellow,~IEEE} 

\thanks{
\update{A preliminary version of this manuscript has been published in the proceedings of the 2023 IEEE Symposium on Security and Privacy (SP)~\cite{deng2023ares}. Qi Li is the corresponding author of this paper.}
}

\thanks{X. Deng, X. Zhao, Z. Liu, Q. Li, M. Xu, K. Xu, and J. Wu are with the Institute for Network Sciences and Cyberspace, Tsinghua University, Beijing 100084, China, e-mail: \{dengxh23@mails., zhuotaoliu@, zhaoy23@mails., qli01@, xumw@, xuke@\}tsinghua.edu.cn, jianping@cernet.edu.cn.}
\thanks{Q. Yin is with the Zhongguancun Laboratory, Beijing 100094, China, e-mail:yinql@zgclab.edu.cn.}
}

\begin{document}

\title{\LARGE \bf Towards Robust Multi-tab Website Fingerprinting}

\maketitle
\begin{abstract}
Website fingerprinting enables an eavesdropper to determine which websites a user is visiting over an encrypted connection. State-of-the-art website fingerprinting (WF) attacks have demonstrated effectiveness even against Tor-protected network traffic. 
However, existing WF attacks have critical limitations on accurately identifying websites in multi-tab browsing sessions, where the holistic pattern of individual websites is no longer preserved, and the number of tabs opened by a client is unknown a priori.
In this paper, we propose \textsf{ARES}, a novel WF framework natively designed for multi-tab WF attacks. 
\update{
\textsf{ARES} formulates the multi-tab attack as a multi-label classification problem and solves it using the novel Transformer-based models. Specifically, \textsf{ARES} extracts local patterns based on multi-level traffic aggregation features and utilizes the improved self-attention mechanism to analyze the correlations between these local patterns, effectively identifying websites.
}
We implement a prototype of \textsf{ARES} and extensively evaluate its effectiveness using our large-scale datasets collected over multiple months. 
The experimental results illustrate that \textsf{ARES} achieves optimal performance in several realistic scenarios. Further, \textsf{ARES} remains robust even against various WF defenses.
\end{abstract}


\section{Introduction}
\label{sec:intro}
Anonymous communication techniques are designed to prevent the content and metadata of network communications from being leaked and/or tampered by malicious activities, such as eavesdropping and man-in-the-middle attack. With millions of daily users~\cite{mani2018understanding}, the Onion Router (Tor) is one of the most popular anonymous communication tools used to protect web browsing privacy. Tor hides user activities by establishing browsing sessions through Tor circuits with randomly selected Tor relays, where data communication in each Tor circuit is encrypted via ephemeral keys and forwarded in fix-sized cells~\cite{dingledine2004tor}. 

Although Tor mitigates the privacy threat to some extent, an adversary can still observe the encrypted traffic of a Tor browsing session and utilize its network traffic patterns (\eg the packet size and interval statistics) to infer the websites visited by the Tor client. 
This technique is referred to as the Website Fingerprinting (WF) attack. 
The rationale behind the WF attack is that the content of each website results in a unique traffic pattern distinguishable from other websites. Prior works~\cite{wang2014effective,panchenko2016website,hayes2016k,abe2016fingerprinting,rimmer2018automated,sirinam2018deep} demonstrated the effectiveness of WF attack, with best attack accuracy exceeding 95\%. In general, these works formulate the WF attack as a classification problem and solve it based on machine learning or deep learning algorithms, such as Support Vector Machine (SVM), Random Forest, and Convolutional Neural Networks (CNN). 

The effectiveness of existing WF attacks relies on a common yet unrealistic assumption. In particular, they assume that \emph{the client only visits a single web page in one browsing session}~\cite{juarez2014critical,wang2016realistically,xu2018multi}. 
This single-page assumption does not always hold in practice since normal clients often open multiple browser tabs simultaneously (or within a very short period)~\cite{juarez2014critical,wang2016realistically,smith2001tcp}. 
A multi-tab browsing session contains the network traffic generated by different web pages such that their patterns are mixed and become more difficult to be identified. 
Prior work~\cite{juarez2014critical} shows that the performance of the traditional WF attacks decreases drastically on multi-tab browsing scenarios. 
To relax this assumption, a series of multi-tab WF attacks have been proposed~\cite{wang2016realistically,xu2018multi,cui2019pse,yin2021automated,guan2021bapm, jin2023tmwf}. 

Most existing multi-tab WF attacks (\eg \cite{wang2016realistically,xu2018multi,cui2019pse,yin2021automated}) share a similar design architecture: they first divide the whole browsing sessions into multiple clean traffic chunks, where each chunk only contains the traffic of a single website, and then infer the visited websites based on each chunk. However, this architecture has three critical drawbacks. 
\first \emph{They require prior knowledge of how many tabs are opened by clients.} Existing multi-tab WF models are trained given a fixed number of tabs, \eg 2 tabs in~\cite{xu2018multi}. Yet, their models are not generic enough to handle other tab numbers. Consequently, these methods often yield very limited accuracy in practice when the number of opened tabs is dynamic and unknown a priori. 
\second Even in such a restricted setting, \emph{these methods are not resilient to the WF defense mechanisms.} WF defenses are designed to perturb the original network traffic patterns by either delaying packet transmissions or padding dummy packets. 
Prior work~\cite{yin2021automated} shows that lightweight WF defenses~\cite{juarez2015wtf, gong2020zero} can significantly limit the effectiveness of existing multi-tab WF attacks. 
\third Further, \emph{their effectiveness further decreases as clients open more browser tabs.} 
The capability of existing multi-tab WF attacks depends on the quality of clean traffic chunks, such as the number of clean chunks and the amount of clean traffic in these chunks. As clients open more browser tabs, it is more difficult to extract clean chunks from a browsing session. 
\update{
Recent studies~\cite{guan2021bapm, jin2023tmwf} explore WF attacks without explicitly dividing the obfuscated traffic into individual chunks. However, these attacks still require prior knowledge of the maximum number of tabs and exhibit significant performance degradation under WF defenses.
}

\noindent \textbf{Our Work.} To address these limitations, we propose a new multi-tab website fingerprinting attack mechanism, \ours.
The core idea of \ours is formulating the multi-tab WF attack as a multi-label classification problem to fundamentally relax the required prior knowledge on the number of tabs opened in a browsing session. 
Towards this end, we design \ours based on a novel multi-tab WF attack framework containing multiple classifiers.
Different from the existing end-to-end WF attacks, we transform the complex multi-label classification problems into the multiple binary classification problem, where
each classifier is responsible for calculating the possibility that whether a specific monitored website is visited. 
Afterwards, \ours regularizes and ranks these possibilities, and then outputs the complete label set for all monitored websites based on a pre-determined threshold. 
Besides the architectural innovation, we also develop a new Transformer model, \transformer, as the robust individual classifier used in \ours, as described below. 

The key observation for designing \transformer is that although a website's clean and holistic traffic pattern is no longer preserved in multi-tab browsing sessions (or simply due to the dummy packets padded by WF defenses), it is still possible to extract multiple \emph{local patterns} for the website from multiple short traffic segments. Thus, \transformer can build signatures for different websites by analyzing the relevance among these local traffic patterns. 
\update{
In its design, \transformer employs a multi-level traffic aggregation module to divide a browsing session into multiple traffic segments, and separately extract packet-based and burst-based aggregation features from these segments. 
These aggregation features effectively capture the robust patterns of different websites within obfuscated traffic. 
}
Then \transformer utilizes a local profiling module to accurately extract the local patterns from aggregation features. Moreover, \transformer designs an improved attention mechanism to further reduce the impact of noises on calculating the relevance among local patterns.  

\xinhao{We extensively evaluate \ours based on large-scale datasets from over 500 thousand multi-tab Tor browsing sessions collected from May 2021 to December 2021 and from June 2022 to November 2022. In addition to multi-tab browsing, we consider
\emph{various real-world complexities in WF attacks}, including \first multiple Tor versions co-exist, \second clients may visit sub-pages beyond the main page in each website, and \third the vantage points for collecting traffic could vary (not just at client-side).}
To the best of our knowledge, our datasets are by far the largest multi-tab WF datasets.

The contributions of our work are three-fold:
\begin{itemize}[leftmargin=*]
    \item We develop \ours, a novel WF attack mechanism specifically designed for the generic multi-tab browsing setting where the number of open tabs 
    \update{is} dynamic and unknown a priori. 
    \item \ours employs a one-vs-all framework containing parallel classifiers to formulate the multi-tab WF attack as a multi-label classification problem. At its core, each classifier is powered by a novel \transformer design that can accurately identify a specific website \emph{without} depending on a clean and holistic traffic pattern from the website.
    \item We implement a prototype of \ours and extensively evaluate it on our large-scale multi-tab browsing datasets. \update{We release the datasets and source code of \ours\footnote{https://github.com/Xinhao-Deng/Website-Fingerprinting-Library}. 
    The experimental results illustrate that \ours effectively achieves the best \textsf{MAP@k} exceeding 0.9. Furthermore, \ours is more resilient against defenses than existing WF attacks and achieves an average performance improvement of 112.74\% over baselines under the realistic WTF-PAD defense.
    }
\end{itemize}

\section{Background}
\label{sec:background}
\subsection{WF Attacks and Defenses}
In general, the \emph{fingerprint} of a website is a combination of network traffic patterns, such as the statistics of packet sizes and intervals when accessing this website. The Website Fingerprinting (WF) attack is a technique that can identify the websites accessed by a client only by analyzing the client's browsing traffic, even in encrypted form. 
When applied by adversaries, the WF attack could compromise normal users' online privacy. 
Yet WF could also assist in crime tracking on the dark web. 

Technically, the WF attack is formulated as a classification problem solvable using machine learning (ML) algorithms.
The existing researches have developed various types of features, \eg the data volume and packet intervals, to profile the encrypted traffic. 
A series of ML-based classifiers (\eg SVM and Random Forest) are used to perform WF attack~\cite{wang2014effective,panchenko2016website,hayes2016k,wang2016realistically,herrmann2009website}. 
In particular, with the emergence of deep learning (DL), DL-based WF attacks achieve automatic feature extraction and higher accuracy~\cite{rimmer2018automated, sirinam2018deep}. 
Further, a study~\cite{sirinam2018deep} shows DL-based WF attacks can effectively bypass the existing WTF-PAD defense~\cite{juarez2015wtf}. However, DL-based WF attacks require a large amount of training data. Sirinam et al.~\cite{sirinam2019triplet} proposed the triple networks based WF attack to solve this problem. 
Still, the above WF attacks assume the client's browsing traffic is purely generated by a single website. The multi-tab attacks~\cite{wang2016realistically,xu2018multi,cui2019pse,yin2021automated} relaxed this assumption. They propose to divide the network traffic to obtain clean traffic chunks to facilitate website fingerprinting. 
\update{
Moreover, the latest multi-tab attacks~\cite{guan2021bapm, jin2023tmwf} leverage Transformer to directly identify obfuscated traffic. However, existing multi-tab WF attacks are not resilient to the WF defenses. Even worse, they require prior knowledge of the number of tabs (or the maximum number of tabs) opened by the user, which is challenging in practice.
}

Website fingerprinting defenses are designed as countermeasures against WF attacks. 
Existing WF defenses mainly fall into three categories: padding-based, mimicry and regularization defense. 
Padding-based defenses (such as WTF-PAD~\cite{juarez2015wtf} and Front~\cite{gong2020zero}) disorder the original traffic pattern by randomly adding dummy packets. Mimicry defenses confuse the traffic pattern, causing the classifiers of WF attacks to falsely identify a website as another one~\cite{nasr2021gan3,panchenko2011website, holland2022regulator}. For example, Decoy~\cite{panchenko2011website} loads a decoy website along with the real website. Regularization defenses make the traffic pattern of all websites fixed by adding dummy packets and delaying packets~\cite{cai2014tamaraw, dyer2012peek}, yet these defenses typically impose high overhead.  

\subsection{Multi-Class and Multi-Label Classification}
In machine learning, the \emph{Multi-Class} classification means that the total number of class labels is greater than two~\cite{wu2004probability} (otherwise, it is a \emph{Binary} classification). 
For example, an adversary has a monitoring set with 100 different websites (\ie class labels) and tries to classify a client's browsing session (\ie an instance) into one of these websites. 

Regardless of the number of class labels, the \emph{Single-Label} classification~\cite{ghamrawi2005collective} only assigns one class label to an instance, \eg classifying the species of an animal. By contrast, the \emph{Multi-Label} classification~\cite{tsoumakas2007multi} may assign one or more class labels to one instance simultaneously. Thus, it is more suitable for the multi-tab web browsing scenario, where each encrypted session contains multiple websites.

\section{Threat Model}
\label{sec:threat_model}

\begin{figure}[t]
  \centering
  \includegraphics[width=0.95\linewidth]{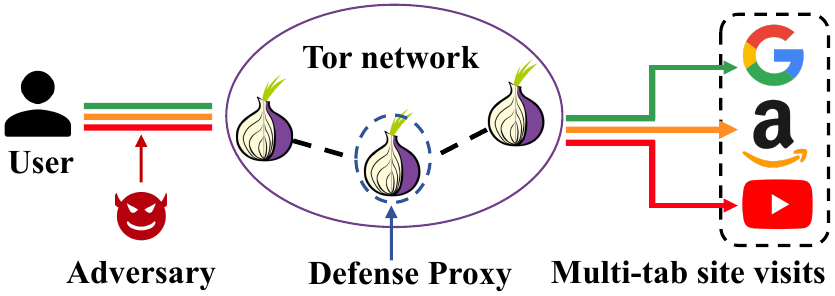}
  \caption{The threat model of \ours. Users open multiple tabs to visit different websites, and the middle nodes of the Tor network may be a defense proxy.}
  \label{fig:model}
\end{figure}

In our threat model, clients access websites using privacy-enhancing techniques like Tor to hide their online activities. 
Each client could open several browser tabs to load multiple pages from different websites simultaneously (or within a short period of time). As a result, a client's browsing session may contain encrypted network packets from multiple websites. Further, the client's browser or on-path Tor relay nodes could have deployed some WF defense mechanisms, such that the traffic patterns of individual websites are no longer preserved. Figure~\ref{fig:model} illustrates our threat model.

We consider a privacy-hungry adversary that primarily focuses on de-anonymizing a client's online activities by inferring the websites visited by the client through website fingerprinting. 
Therefore, the adversary could deploy multiple traffic mirroring points to record the client's encrypted network traffic, even before the traffic enters the entry node of the Tor network. Yet, actively delaying or even discarding the client's network traffic is out of the scope of our threat model.

Compared with the original multi-tab WF threat models~\cite{wang2016realistically,xu2018multi,cui2019pse,yin2021automated,guan2021bapm}, our model is more realistic, yet more challenging, for the following three reasons. 
\first \emph{We consider that the client could have deployed existing WF defenses.} As a result, the traffic pattern of individual websites could have been perturbed by these anti-WF techniques. 
\second \emph{We consider that the number of tabs opened by the client is dynamic and unknown a priori}. Prior WF mechanisms assume that the clients always open a fixed number of tabs (\eg two tabs in~\cite{yin2021automated}) since their models have to be trained and tested under the same specific setting. This is restrictive and unrealistic.
\third \emph{We consider critical real-world complexities in WF attacks}. 
Existing WF attacks~\cite{hayes2016k,rimmer2018automated,sirinam2018deep,sirinam2019triplet,wang2020high} are evaluated using over-simplified scenarios, where clients use the same version of Tor Browser,  clients only browse the homepage of websites, and network traffic is collectible at the client-side, etc. These assumptions are largely incorrect in practice. 
Therefore, our design considers a more practical threat model, where multiple versions of Tor Browsers can co-exist, clients can visit the sub-pages of websites, and different vantage points for traffic collection other than at the client side are evaluated. 

Similar to existing arts~\cite{hayes2016k,rimmer2018automated,sirinam2018deep,sirinam2019triplet}, our model contains two attack scenarios: closed-world and open-world.
The closed-world scenario assumes that clients will only visit a small set of websites (\eg the Alexa Top 100 websites). 
In this case, the adversary has the resources to collect data from all these websites (referred to as \emph{monitored websites}). In the open-world scenario, clients can visit arbitrary websites, and therefore the adversary may only possess training data for a subset of the websites.
\section{Design of \ours}
\label{sec:design}

\begin{figure*}[t]
  \centering
  \includegraphics[width=\linewidth]{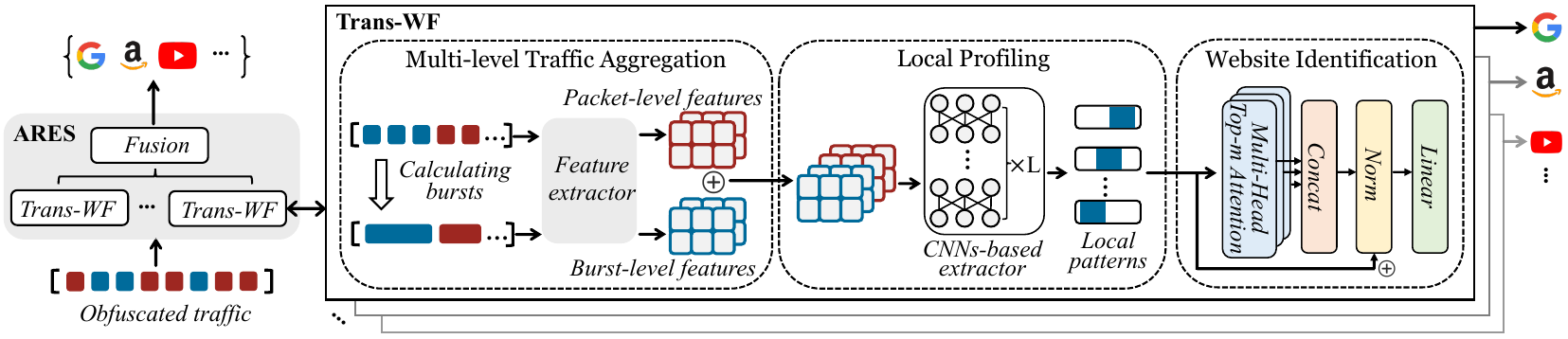}
  \caption{The overview of \ours.}
  \label{fig:overview}
\end{figure*}

In this section, we present the design detail of \ours. We start with an overview of \ours before delving into its individual components. 

\subsection{Overview}

As discussed in Section~\ref{sec:intro}, prior multi-tab WF attacks require prior knowledge of the number of tabs opened in a browsing session. To fundamentally relax this limitation, \ours regards the multi-tab attack as a multi-label classification problem. 

It is challenging to solve the multi-label classification problem because the traffic from different websites is mixed together and the number of visited websites is unknown and dynamic.
\update{In particular, due to the high-dimensional features, mixed website traffic, and noises generated by WF defenses, the performance of existing WF attacks degrades significantly.}
Therefore, \ours builds a multi-tab WF attack framework with multiple classifiers, and
each classifier is utilized to calculate the possibility of whether a specific website is accessed.
Then, \ours integrates the results of individual classifiers to generate the complete label set for all monitored websites without prior knowledge of the number of tabs. 
Moreover, we develop a novel Transformer~\cite{vaswani2017attention} model called Transformer for Website Fingerprinting (\transformer), as the classifier used in \ours. The design of \transformer is based on a key observation that a website's local patterns are still extractable from multiple short traffic segments, even when the entire traffic pattern is no longer preserved in a multi-tab browsing session and under defenses. 
Thus, \transformer can build robust signatures for different websites based on these local patterns. 

\update{
In Figure 2, we illustrate the architecture of \ours. At a high level, \ours consists of $N$ \transformer, where N represents the number of monitored websites, and the i-th \transformer is used to identify the i-th monitored website. \ours fuses the outputs of all \transformer models to output a label set for all monitored websites. The \transformer model consists of three modules designed to robustly identify obfuscated traffic of the specific website, including multi-level traffic aggregation, local profiling, and website identification.

\noindent \textbf{Multi-level Traffic Aggregation.} 
The multi-level traffic aggregation module extracts features containing local website information from obfuscated traffic. Although global website information within the obfuscated traffic is disrupted, sufficient local website information is retained within its sub-segments. Therefore, \transformer first divides the traffic and then extracts packet-level and burst-level aggregation features from each segment to preserve local website information. The details of this module will be described in Section~\ref{sec:extract-agg}.

\noindent \textbf{Local Profiling.} The local profiling module utilizes a convolutional neural network (CNN) to extract local traffic patterns that represent key elements of the specific website. The obfuscated traffic generated by multi-tab browsing is dynamic, resulting in local traffic patterns having variable positions within the traffic. By leveraging the translation invariance of CNNs, \transformer can effectively extract local traffic patterns that appear in any position, thus supporting website identification. We will describe this module in Section~\ref{sec:pattern-extra}.

\noindent \textbf{Website Identification.} The website identification module robustly identifies obfuscated traffic through an improved self-attention mechanism. Noise packets generated by multi-tab browsing and WF defenses pose significant challenges for website identification. However, there are correlations among different local patterns within obfuscated traffic. \transformer utilizes the top-m self-attention mechanism that mitigates the interference of noise packets and effectively analyzes correlations between local patterns. We will present the details of website identification in Section~\ref{sec:web-ident}.
}

\begin{figure}[t]
  \centering
  \includegraphics[width=0.8\linewidth]{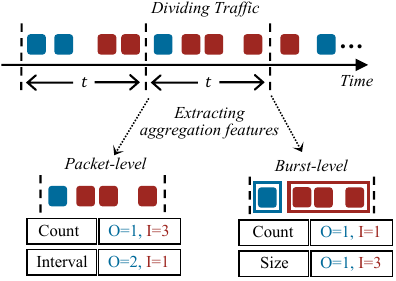}
  \caption{Dividing obfuscated traffic and extracting multi-level traffic aggregation features involving packet-level and burst-level features.}
  \label{fig:traffic-agg}
\end{figure}

\update{
\subsection{Multi-level Traffic Aggregation}
\label{sec:extract-agg}

The multi-level traffic aggregation module aims to extract traffic features containing website information from multi-tab obfuscated traffic while maintaining robustness even under WF defenses.
Extracting effective website features from obfuscated traffic is challenging because noise packets introduced by multi-tab browsing or WF defenses disrupt the global traffic patterns of websites. 
The motivation behind this module is that sub-segments of obfuscated traffic still contain sufficient local patterns associated with key elements of the website. 
Therefore, the multi-level traffic aggregation module focuses on extracting meaningful local features from short traffic segments. These local features preserve sufficient information to characterize individual websites while being resilient to the noise and variability introduced by multi-tab browsing and WF defenses.

As shown in Figure~\ref{fig:traffic-agg}, we illustrate the process of multi-level traffic aggregation. 
We extend the traffic aggregation used in existing single-tab WF attacks~\cite{shen2023RF, deng2024ccs}. Specifically, 
\transformer first divides the obfuscated traffic into fixed-length sub-segments based on a uniform time interval $t$.
Each subsegment contains local traffic patterns that are relevant and consistent to the website.
Through traffic division, this module mitigates the impact of global noise and facilitates further feature extraction. Next, \transformer extracts aggregation features separately for incoming and outgoing traffic from each sub-segment, including packet-level features and burst-level features.
The packet-level features include the total number of packets as well as the average interval between consecutive packets.  
These features capture fine-grained packet behavior and temporal dynamics, which are unique to specific websites as they are associated with the key elements of those websites.
Unlike packet-level features, burst-level features are coarse-grained representations of traffic transmission. A burst represents a sequence of consecutive packets in the same direction. 
The burst-level features extracted by \transformer include the total number of bursts and the average burst size, effectively capturing higher-level traffic patterns that remain stable even under obfuscation. 
In summary, the traffic aggregation features of each segment consist of 8 values, representing the key patterns of incoming and outgoing traffic within the sub-segment. These aggregated features from all segments are concatenated and fed into the local profiling module.

This multi-level traffic aggregation provides several advantages. 
Traffic division minimizes the impact of disrupted global traffic patterns by isolating meaningful patterns within shorter sub-segments.
By extracting features at both packet and burst levels, \transformer ensures a comprehensive representation of website patterns, capturing both fine and coarse-grained behaviors. 
Moreover, these features exhibit strong robustness against noise and temporal perturbations introduced by WF defenses and multi-tab browsing, enabling the effective identification of highly obfuscated traffic.
In practice, this module achieves a balance between feature complexity and computational efficiency. The selected features are carefully designed to provide sufficiently robust representations, and we show the ablation analysis of the aggregation features in Section~\ref{sec:ablation}.

Overall, the multi-level traffic aggregation module enables \transformer to extract accurate and resilient website fingerprints from multi-tab browsing sessions. It isolates and extracts robust local features from obfuscated traffic, ensuring that our attack remains effective even under complex real-world conditions.
}

\subsection{Profiling Local Patterns}
\label{sec:pattern-extra}

\begin{figure}[t]
  \centering
  \includegraphics[width=\linewidth]{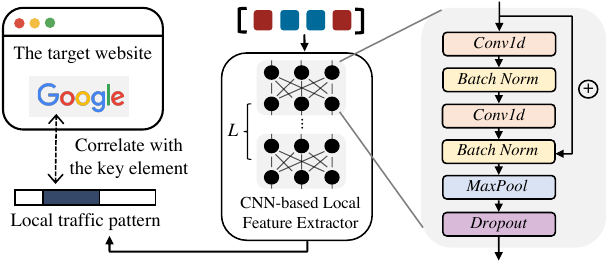}
  \caption{Profiling the traffic pattern generated from each website. The local traffic pattern is correlated with the key elements in a website, which can be extracted by CNN due to its invariant translation.}
  \label{fig:local_profiling}
\end{figure}

The local profiling module is applied to profile the local patterns of a monitored website by extracting the local feature vectors from \update{multi-level aggregation features}. 
This is challenging for the following two reasons:
\first the locations of the packet sequences representing different local patterns are not fixed; \second the irreverent packets in the same segment generated from other websites or WF defenses create non-trivial noises.  
To overcome this challenge, we design our local profiling module based on Convolutional Neural Networks (CNN). 
CNN has the characteristic of invariant translation~\cite{kayhan2020translation}, \ie it can profile the input data into the same embedding vectors regardless of how the input data is shifted. Moreover, prior WF attacks have demonstrated that CNN is more resilient against noises~\cite{sirinam2018deep,rahman2019tik}. 

As shown in Figure~\ref{fig:local_profiling}, the local profiling module contains $L$ blocks and each block consists of two one-dimensional convolution layers (Conv1d), two batch normalization layers (BN) with the ReLU activation function (ReLU) and a max-pooling layer. Besides, we introduce two additional regularization techniques to further enhance our module. \first \textit{Residual connection.} It propagates the intermediate output of lower layers to higher layers through skip connections to prevent gradient vanishing. \second \textit{Dropout.} It randomly drops some units (along with their connections) from the neural network during training to alleviate over-fitting.

In each block, the input is first fed into two convolution layers and two batch normalization layers, to extract the local features. These local feature vectors (\ie the output of the last batch normalization layer) are connected with their original input via the residual connection, and then they will be fed into the max-pooling layer, for the purpose of retaining the most representative features while progressively reducing their sizes. Thus, the small perturbations in the input traffic segments can be filtered by the max-pooling layer. 


\subsection{Identifying Websites}
\label{sec:web-ident}

The website identification module is in charge of analyzing the relevance among local patterns to identify whether a monitored website is visited in the multi-tab browsing session. 
The self-attention mechanism proposed in the transformer model~\cite{vaswani2017attention} is a reasonable choice for this goal. 
The self-attention mechanism is widely applied in natural language processing and computer vision~\cite{vaswani2017attention,han2021transformer,wang2021pyramid,liu2021swin}, which can capture the dependencies within a sequence.
Therefore, the self-attention mechanism can effectively analyze the dependencies of multiple local patterns, and thus identify the target website.
Since the number of tabs opened by the client is dynamic, we use a multi-head self-attention mechanism to capture the information of the target website under the different numbers of tabs.
Furthermore, we design the top-\emph{m} attention, an improved self-attention mechanism, to enhance the model robustness under WF defenses.

The attention mechanism is a function that computes the relevance between a \emph{query} and a set of \emph{key-value} pairs, where the \emph{query}, \emph{key}, and \emph{value} are all vectors projected from the input data individually~\cite{vaswani2017attention}. In particular, the attention function first calculates the weight of each \emph{value} using a compatibility function of the \emph{query} 
and its corresponding \emph{key}, and then produces a weighted sum of all \emph{values} as the output that represents the relevance between the \emph{query} and \emph{key-value} pairs. When we apply this mechanism to correlate different segments of the same sequence, namely the self-attention~\cite{vaswani2017attention}, it can convert the sequence into a new representation that reveals its internal relevance. 
Thus, we can take the local feature vectors as the input of the self-attention function, and utilize the corresponding output as the fingerprint of a monitored website. 

We illustrate this procedure using the vanilla attention mechanism~\cite{vaswani2017attention} at first. Let $\mathbf{Q}$, $\mathbf{K}$, and $\mathbf{V}$ donate the \emph{query}, \emph{key}, and \emph{value} matrices, respectively. As shown in Equation~(\ref{equ:qkv}), these matrices can be achieved via linear projections of a batch of input data $\bm{X}$ (\ie the local feature vectors), where $\bm{X}\in \mathbb{R}^{b \times d_m}$, $b$ is the number of local features (\ie batch size) and $d_m$ represents the dimension of a local feature:
\begin{equation}
    \bm{Q=XW}^Q, ~ \bm{K=XW}^K, ~ \bm{V=XW}^V, 
	\label{equ:qkv}
\end{equation}
where $\bm{W}^Q, \bm{W}^K, \bm{W}^V \in \mathbb{R}^{d_m \times d}$ are the matrices for projections and $d$ is the dimension of an output vector. Note that these projection matrices will be learned during model training. Then, the output of this attention function can be computed via Equation~(\ref{equ:attn-1}):
\begin{equation}
    \textsf{Attention}(\bm{Q,K,V}) = \textsf{softmax}(\frac{\bm{Q} \bm{K}^T}{\sqrt{d}}) \bm{V}. 
	\label{equ:attn-1}
\end{equation}
In general, this equation computes the dot products of each query with all keys, scales these results by dividing $\sqrt{d}$, and applies a \textsf{softmax} function to obtain the weights of each value.


However, when identifying a monitored website under multi-tab browsing scenarios, the vanilla self-attention mechanism has a severe shortcoming in that it is not resilient to the traffic noises generated by other websites and WF defenses. In particular, this mechanism contains a fully-connected attention layer such that the output vector for an input vector (\ie a local feature vector) depends on the relevance between this input with all other inputs (\ie all other local feature vectors). As a result, the local features extracted from noisy traffic inevitably reduce the accuracy of the output. 

\begin{figure}[t]
  \centering
  \includegraphics[width=\linewidth]{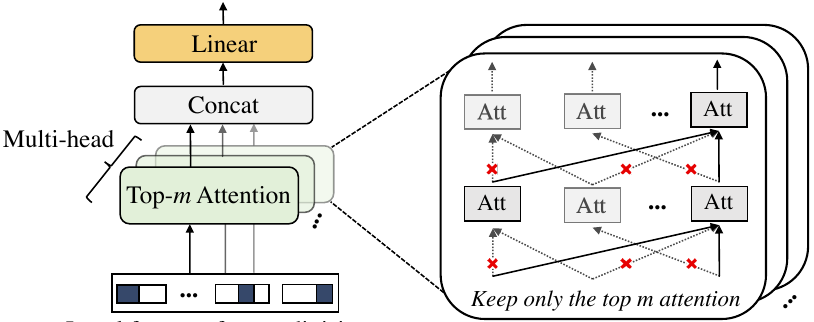}
  \caption{The multi-head top-m attention method correlates local traffic patterns for website fingerprinting even in the presence of noise interference. Different from the full connection of the original Transformer, \transformer keeps only the top-m attention.}
  \label{fig:topm-atten}
\end{figure}

To handle this issue, we design an improved attention layer, namely top-\emph{m} attention, based on~\cite{wang2021kvt}. This layer calculates the output for an input vector based on the top-\emph{m} weight values computed by its corresponding \emph{query} and all \emph{keys}, rather than all weight values. In general, the traffic of the monitored website is less correlated with the traffic generated by other websites or WF defenses than itself.  
This means that the monitored website’s local features and the local features from other websites tend to have smaller attention-based weight values. Thus, we can filter out the interference from the traffic noises via the top-\emph{m} selection strategy. 
Let $\mathbf{Q}$, $\mathbf{K}$, and $\mathbf{V}$ donate the \emph{query}, \emph{key}, and \emph{value} matrices, respectively.
We formally describe this new attention layer design in Equation~(\ref{equ:attn-2}):
\begin{equation}
    Attention^{Top-m}(\bm{Q,K,V}) = softmax(\Gamma(\frac{\bm{Q} \bm{K}^T}{\sqrt{d}})) \bm{V}, 
	\label{equ:attn-2}
\end{equation}
\begin{equation}
    [\Gamma(A)]_{ij} =
    \begin{cases}
    A_{ij},  & \text{$A_{ij}$ is the top-\emph{m} largest elements in row j},\\
    \epsilon, & \text{otherwise}, 
    \end{cases}
	\label{equ:top-m}
\end{equation}
where $\Gamma(\cdot)$ defines a row-wise top-\emph{m} selection operation and $\epsilon$ is a small enough constant. In our website identification module, we replace the vanilla attention layer with our new design. 

\xinhao{
As the number of tabs opened by the client is unknown and dynamic, the correlation between local patterns varies with the number of tabs. 
Therefore,}
we parallel multiple top-\emph{m} attention layers to compose a \emph{multi-head top-m attention layer}. \xinhao{As shown in Figure~\ref{fig:topm-atten}}, it allows \transformer to jointly capture the relevance among the local features 
\xinhao{even in the dynamic number of tabs}
, such that the relevance representations can be enriched to achieve even more accurate website identifications. For the i-th head, its output is computed via Equation~(\ref{equ:multi-head}):
\begin{equation}
    head_i =  Attention^{Top-m}(\bm{QW}_i^Q, \bm{KW}_i^K, \bm{VW}^V_i),
	\label{equ:multi-head}
\end{equation}
where $\bm{W}_i^Q, \bm{W}_i^K, \bm{W}_i^V \in \mathbb{R}^{d \times d_h}$ are the weight matrices specific to this head, where $d_h$ is the dimension of the output vector of each head. 
Let $h$ denotes the number of heads, and we set $d_h=d/h$.
Note that each head performs its own task individually. Then, the results of each head will be concatenated and transformed by a \xinhao{linear} projection. 
Let $\Lambda(\bm{X})$ denote the output of our multi-head top-\emph{m} attention layer.
Finally, we can produce $\Lambda(\bm{X})$ via Equation~(\ref{equ:head-output}).
\begin{equation}
    \Lambda(\bm{X}) = \textsf{Concat}(head_1,\ldots,head_h)\bm{W}^O,
	\label{equ:head-output}
\end{equation}
where $\bm{W}^O \in \mathbb{R}^{hd_h \times d}$ is the weight matrix.
With the output of the attention layer, we utilize a batch normalization layer and a Multilayer perceptron (MLP) to identify the existence of a target website. Also, we apply the techniques of residual connection and dropout to avoid the problems of gradient vanishing and over-fitting, respectively. The identification result $\Phi(\bm{X})$ of a target website can be computed as follows:
\begin{equation}
    \Phi(\bm{X}) = \bm{MLP}(\bm{LN}(\bm{X} + Dropout(\Lambda(\bm{X})))),
	\label{equ:norm}
\end{equation}
\begin{equation}
    \bm{LN}(\bm{X}) = \frac{\bm{g}}{\sqrt{\sigma^2+\epsilon}} \odot (\bm{X}-\mu) + \bm{b} ,
	\label{equ:ln}
\end{equation}
where $\bm{LN}$ is the layer normalization, $\bm{g,b}$ are the gain and bias parameters, $\mu,\sigma$ are the mean and the variance of $\bm{X}$, $\odot$ is the element-wise multiplication between two vectors, and $\epsilon$ is a small constant to prevent division by zero. The MLP utilizes the common softmax function.

To mitigate the potential over-fitting of \transformer,
we thus use \textit{Droppath}~\cite{huang2016droppath} in \transformer.
The \textit{Droppath} randomly drops some training instances in the residual connection during training, causing these instances to skip part of the training. 
In particular, the \textit{Droppath} achieves differential model training, which can alleviate the over-fitting.


\section{Evaluation} 
\label{sec:evaluation}

In this section, we evaluate the effectiveness of \ours with real-world multi-tab datasets. We also compare the performance of our work with the state-of-the-art WF attacks.

\subsection{Experimental Setup}
\label{sec:experimental_setup}

\begin{table}[t]
\footnotesize
\centering
\caption{\update{Parameter settings 
in our evaluation.}}
\label{tab:parameters}
\begin{tabular}{ccc}
\toprule
\textbf{Module Part} & \textbf{Design Details} & \textbf{Value} \\ \midrule
\multirow{2}{*}{Traffic Aggregation} 
 & Input dimension $d$ & 8000 \\
 & Interval $t$ & 20 ms \\
 \midrule
\multirow{4}{*}{Local Profiling}
 & Number of blocks & 4 \\
 & Kernel size & 7 \\
 & Pool size & 8 \\
 & Output dimension & {256} \\ \midrule
\multirow{3}{*}{Website Identification} 
 & Number of heads & 2 \\
 & Number of layers $n$ & 4 \\
 & Value of $m$ & 20 \\
\bottomrule
\end{tabular}
\end{table}

\noindent \textbf{Implementation}. We prototype \ours using PyTorch with over 1,500 lines of code~\cite{deng2024ccs}. 
\update{
Table~\ref{tab:parameters} presents the default parameter values of \ours, and we further study the impact of parameter choice in Section~\ref{sec:parameter_analysis}.
Note that the original \ours requires an individual \transformer model for each monitored website, resulting in significant overhead during training and deployment. To address this challenge, we reduce the number of required \transformer models by optimizing the multi-label one-versus-all loss based on maximum entropy~\cite{he2021multi}, thereby improving the practicality of \ours for real-world applications.
}

\noindent \textbf{Datasets}. We develop an automated Tor browsing tool with over 1,000 lines of code (LOC) based on the Tor Browser and Selenium framework~\cite{Selenium}. We deploy our tool on 40 different cloud servers located in different regions to simulate Tor clients located across the globe. Our data collection is divided into two phases from May 2021 to December 2021 and from June 2022 to November 2022.
\update{
We utilize various methods to filter out noise traffic and improve the quality of datasets. 
For example, We enhanced datasets by employing a ResNet-based image classification model to filter failed page loads.
Further details of the dataset construction can be found in~\cite{deng2023ares}. Our datasets comprise seven categories of data.

\begin{itemize}[leftmargin=*]
\item {\textbf{Closed-world multi-tab dataset}}: 
We selected the Alexa top 100 websites as monitored websites and collected multi-tab browsing traffic for different website combinations. The dataset contains over 230,000 instances of obfuscated traffic with the number of tabs ranging from 2 to 5.

\item {\textbf{Open-world multi-tab dataset}}:
In addition to the 100 monitored websites, we randomly selected websites from the Alexa Top 20,000 websites as non-monitored websites and collected over 250,000 instances of multi-tab obfuscated traffic involving simultaneous browsing of monitored and non-monitored websites, with the number of tabs ranging from 2 to 5.

\item {\textbf{Dataset with Random defense}}: Randomly padding dummy packets is a common defense strategy to minimize the data overhead of the defense. This dataset contains over 50,000 instances of 2-tab obfuscated traffic with the Random defense.

\item {\textbf{Dataset with WTF-PAD defense}}: The WTF-PAD~\cite{juarez2015wtf} defense effectively disrupts traffic patterns through adaptive padding with dummy packets. A circuit-level variant of the WTF-PAD defense has been deployed in Tor~\cite{circuitpadding}. This dataset contains over 50,000 instances of 2-tab obfuscated traffic with the WTF-PAD defense.

\item {\textbf{Dataset with Front defense}}: The Front~\cite{gong2020zero} defense generates insertion times for dummy packets based on the Rayleigh distribution, aiming to obscure website information contained in the front of traffic. This dataset contains over 50,000 instances of 2-tab obfuscated traffic with the Front defense.

\item {\textbf{Dataset with RegulaTor defense}}: The RegulaTor~\cite{holland2022regulator} defense combines dummy packet padding and packet delays to obscure burst patterns in the traffic. It employs distinct strategies for upload and download traffic. This dataset contains over 50,000 instances of 2-tab obfuscated traffic with the RegulaTor defense.

\item {\textbf{Dataset with Dynamic Settings}}: In realistic scenarios, the number of tabs and the deployed defenses are unknown and dynamic. Therefore, we randomly sample and combine traffic from different numbers of tabs to create a dataset with the dynamic number of tabs. Furthermore, we randomly sample and combine obfuscated traffic with four defenses to construct a dataset with dynamic defenses. This dataset contains over 100,000 instances.

\end{itemize}

Stronger WF defenses have been studied~\cite{cai2014tamaraw, gong2022surakav, shen2024real}, but these defenses are not practically deployable due to the excessive overhead. High overhead could cause functionality issues in Tor relay nodes. Therefore, following previous attacks~\cite{sirinam2018deep, shen2023RF, deng2024ccs}, we select four representative defenses.
}

\noindent \textbf{Baselines}. We use seven representative WF attacks as our baseline methods, divided into three categories.

\begin{itemize}[leftmargin=*]
\item {\textbf{Single-tab WF attacks}}: \update{We select two classical single-tab WF attacks: \varcnn~\cite{bhat2019var} and \netclr~\cite{bahramali2023netclr}. \varcnn leverages deep learning to automatically extract website fingerprints. \netclr incorporates data augmentation and contrastive learning to enable more realistic WF attacks.}

\item {\textbf{WF attacks resilient to defenses}}: \update{
We select three state-of-the-art (SOTA) single-tab WF attacks that are resilient to WF defenses: \df~\cite{sirinam2018deep}, \tiktok~\cite{rahman2019tik}, and \rf~\cite{shen2023RF}. To bypass WF defense, \df proposes more sophisticated DL-based models. \tiktok and \rf extract features based on timing and aggregation information, respectively, significantly enhancing the robustness of WF attacks.
}

\item {\textbf{Multi-tab WF attacks}}: \update{
We choose two state-of-the-art (SOTA) multi-tab WF attacks: \bapm~\cite{guan2021bapm} and \tmwf~\cite{jin2023tmwf}.
\bapm and \tmwf effectively utilize self-attention mechanisms to effectively identify multi-tab obfuscated traffic.
}
\end{itemize}

Note that we follow prior works~\cite{guan2021bapm} to extend the baselines for adaptation to multi-tab WF attacks. Specifically, we replace the loss of three single-tab WF attacks with the binary cross-entropy loss and use the sigmoid layer as their output layer to optimize the model training. This is a minor extension of models and enables multi-tab WF attacks.
\update{
Particularly, we replace the adaptive pooling layer in the \rf model with a linear layer. The adaptive pooling layer mixes information from all websites in the obfuscated traffic, preventing the \rf model from being successfully trained.
Furthermore, \tmwf and \bapm require prior knowledge of the maximum number of tabs that the user browses. To eliminate this dependency, we fuse the model's predictions across all tabs and train the model using binary cross-entropy loss.
}

\noindent \textbf{Metrics}. 
We use three widely-used multi-label classification metrics: \textsf{AUC}~\cite{ling2003auc}, \textsf{P@K} and \textsf{MAP@K}~\cite{liu2017deep}. These metrics evaluate the predicted label set of each instance individually so that we can calculate the average results for all testing instances. 
Recall that $\mathbf{y}$ is the true label vector for an instance $\bm{x}$ and if $\bm{x}$ browses the i-th website, then $\mathbf{y}_i = 1$. Otherwise, $\mathbf{y}_i = 0$. For $\bm{x}$, $\hat{y}$ indicates the predicted label vector (\ie the probability of each website). 
\textsf{P@K} and \textsf{MAP@K} are two metrics for measuring the \emph{k} websites with top-k highest probabilities in $\hat{y}$.
In particular, \textsf{P@K} measures how many browsed websites existed in the top-\emph{k} predicted websites. 

We calculate \textsf{P@K} for $\bm{x}$ via Equation~(\ref{equ:p-k}), where $r_k(\hat{y})$ is the set of websites with top-\emph{k} highest probabilities in $\hat{y}$. 

\begin{equation}
    \textsf{P@k} = \frac{1}{k} \sum_{l \in r_k(\hat{y})} \mathbf{y}_l.
    \label{equ:p-k}
\end{equation}

The \textsf{MAP@K} metric extends \textsf{P@K}, to further evaluate whether the browsed websites have higher probabilities than the non-browsed websites in the top-\emph{k} prediction result. Since a \textsf{MAP@K} score integrates the \textsf{P@K} scores with different \emph{k} values, it is not necessary to change the \emph{k} value for a specific tab setting. We can compute \textsf{MAP@K} as follows: 
according to Equation~(\ref{equ:map-k}).

\begin{equation}
    \textsf{MAP@k} = \frac{\sum_{i=1}^k P@i}{k}.
    \label{equ:map-k}
\end{equation}

Furthermore, we use two metrics, \ie \textsf{Precision} and \textsf{Recall} to evaluate the prediction results for each website. 
Based on prediction results, we can calculate the numbers of true positive instances (TP), false positive instances (FP), true negative instances (TN), and false negative instances (FN) for each website, respectively. These three metrics for each website can be computed as: $\textsf{Precision} = \frac{TP}{TP+FP}$, and $\textsf{Recall} = \frac{TP}{TP+FN}$. We can compute the average results for all websites. 

\begin{table*}[t]
\centering
\small
\caption{\update{Comparisons with prior arts with the multi-tab obfuscated traffic in the closed-world scenario.}}
\label{tab:closed-auc}
\begin{tabular}{c|ccc|ccc|ccc|ccc} \toprule

& \multicolumn{3}{c|}{\textbf{2-tab}}
& \multicolumn{3}{c|}{\textbf{3-tab}} 
& \multicolumn{3}{c|}{\textbf{4-tab}} 
& \multicolumn{3}{c}{\textbf{5-tab}} \\ 
 \cmidrule(lr){2-4} 
 \cmidrule(lr){5-7} 
 \cmidrule(lr){8-10} 
 \cmidrule(lr){11-13}  
\multirow{-2}{*}{} & 
\multicolumn{1}{c}{\textbf{AUC}} & \multicolumn{1}{c}{\textbf{P@2}} & \multicolumn{1}{c|}{\textbf{MAP@2}} & 
\multicolumn{1}{c}{\textbf{AUC}} & \multicolumn{1}{c}{\textbf{P@3}} & \multicolumn{1}{c|}{\textbf{MAP@3}} & 
\multicolumn{1}{c}{\textbf{AUC}} & \multicolumn{1}{c}{\textbf{P@4}} & \multicolumn{1}{c|}{\textbf{MAP@4}} & 
\multicolumn{1}{c}{\textbf{AUC}} & \multicolumn{1}{c}{\textbf{P@5}} & \multicolumn{1}{c}{\textbf{MAP@5}} \\  

\midrule 

\netclr 
& {0.849} & {0.345} & {0.415} 
& {0.740} & {0.211} & {0.269} 
& {0.698} & {0.187} & {0.235} 
& {0.655} & {0.161} & {0.196} 
\\
\bapm
& {0.935} & {0.528} & {0.622} 
& {0.872} & {0.384} & {0.492} 
& {0.839} & {0.354} & {0.452} 
& {0.800} & {0.307} & {0.393} 
\\
\df
& {0.944} & {0.601} & {0.712} 
& {0.864} & {0.424} & {0.566} 
& {0.830} & {0.374} & {0.510} 
& {0.776} & {0.308} & {0.434} 
\\
\rf
& {0.948} & {0.641} & {0.729} 
& {0.880} & {0.471} & {0.596} 
& {0.859} & {0.434} & {0.561} 
& {0.798} & {0.338} & {0.446} 
\\
\tiktok
& {0.958} & {0.647} & {0.754} 
& {0.876} & {0.451} & {0.597} 
& {0.839} & {0.387} & {0.529} 
& {0.781} & {0.311} & {0.438} 
\\
\varcnn
& {0.961} & {0.655} & {0.752} 
& {0.906} & {0.512} & {0.654} 
& {0.881} & {0.471} & {0.605} 
& {0.842} & {0.394} & {0.522} 
\\
\tmwf
& {0.972} & {0.722} & {0.788} 
& {0.946} & {0.635} & {0.720} 
& {0.931} & {0.607} & {0.685} 
& {0.893} & {0.500} & {0.586} 

\\
\ours
& \textbf{0.995} & \textbf{0.904} & \textbf{0.938} 
& \textbf{0.990} & \textbf{0.873} & \textbf{0.916} 
& \textbf{0.989} & \textbf{0.884} & \textbf{0.922} 
& \textbf{0.988} & \textbf{0.869} & \textbf{0.909} 

\\
\bottomrule
\end{tabular}
\end{table*}
\begin{figure*}[t]
  \centering
  \includegraphics[width=\linewidth]{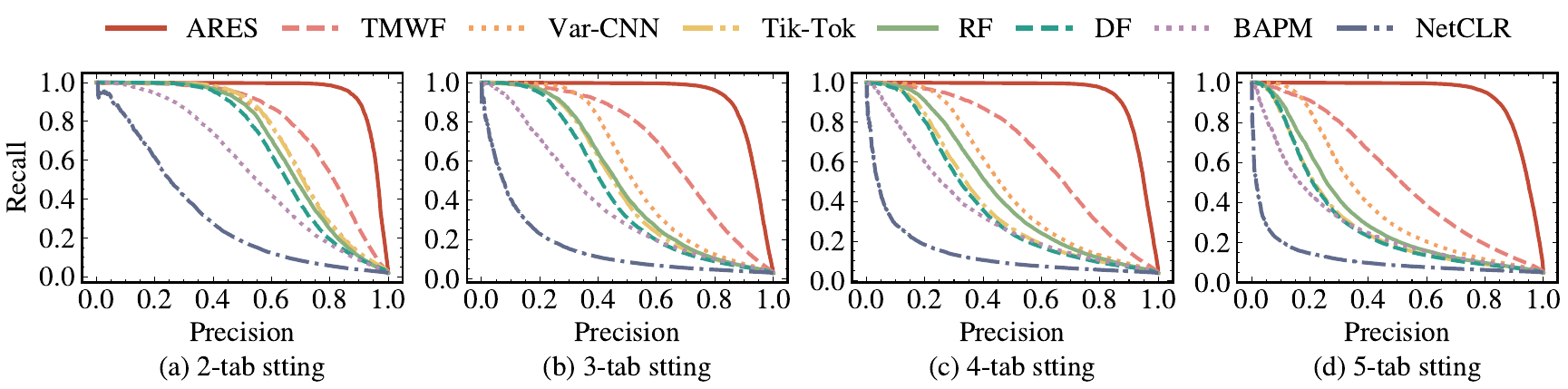}
  \caption{Precision-recall curves of multi-tab WF attacks in the closed-world scenario.}
  \label{fig:closed_pr_curve}
\end{figure*}

\subsection{Multi-tab WF Attacks in the Closed-World.}
\label{sec:multi-tab-attack}

We first evaluate \ours in the closed-world scenario. 
\update{Table~\ref{tab:closed-auc} shows the \textsf{AUC}, \textsf{P@k}, \textsf{MAP@k} results for multi-tab WF attacks. \ours achieves the best performance across different multi-tab settings. Specifically, \ours achieves a \textsf{P@2} of 0.904 in the 2-tab setting, outperforming \netclr, \bapm, \df, \df, \tiktok, \varcnn, and \tmwf, whose \textsf{P@2} are 0.345, 0.528, 0.601, 0.641, 0.647, 0.655, and 0.722, respectively. 
Even in the most challenging 5-tab setting, \ours achieves a \textsf{P@5} of 0.869 and a \textsf{MAP@5} of 0.909, representing an average improvement of 190.83\% and 135.02\% over the baselines, respectively. 
We observe that as the number of tabs increases, the performance of existing attacks declines significantly, whereas \ours maintains greater stability. Compared to the baselines, \ours exhibits higher prediction probabilities for visited websites, effectively identifying all the visited sites within obfuscated traffic. 

We further evaluate multi-tab attacks using the \textsf{AUC} metric. The results show that when the number of tabs is 2, 3, 4, and 5, \ours achieves an average \textsf{AUC} improvement of 6.25\%, 14.48\%, 18.6\%, and 25.71\% over the baselines, respectively. These findings demonstrate that \ours can more effectively differentiate obfuscated traffic from various websites, reducing misidentifications. Furthermore, as the number of tabs increases, the performance advantage of \ours over the baselines becomes more obvious.

Figure~\ref{fig:closed_pr_curve} illustrates the precision-recall curves of all WF attacks under different multi-tab settings. The precision-recall curve shows the average precision and recall of identifying all websites at various thresholds, with curves closer to the upper-right corner indicating better attack performance. \ours achieves the best precision-recall curves, demonstrating its optimal identification precision and recall for multi-tab obfuscated traffic.
}

\noindent \textbf{Remark.} Overall, the above evaluation results demonstrate that \ours can accurately identify the browsed websites in multi-tab scenarios. Moreover, it validates the performance advantage of our work over all existing WF attacks, especially when clients open more tabs.
The main reason is that \ours can build robust signatures for different websites based on the local features extracted from obfuscated traffic. 

\begin{figure}[t]
  \centering
  \includegraphics[width=\linewidth]{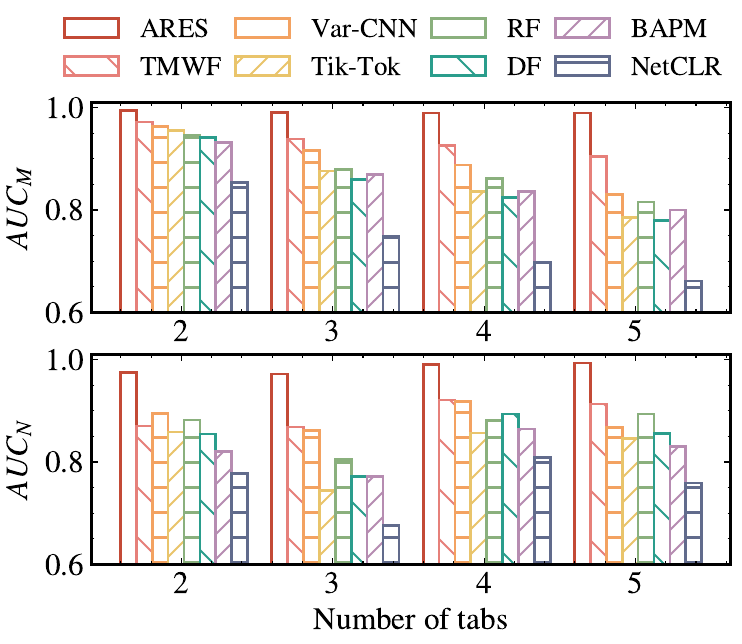}
  \caption{Comparison of AUC of monitored websites $\textsf{AUC}_M$ and non-monitored websites $\textsf{AUC}_N$ with different tab settings in the open-world scenario.}
  \label{fig:open_auc}
\end{figure}


\begin{table}[t]
\small
\centering
\caption{\update{Comparison of \textsf{MAP@K} for multi-tab WF attacks under different tab settings in the open-world scenario.}} 
\label{tab:open-mapk}
\centering
\begin{tabular}{c|cccc}  \toprule
 & \textbf{MAP@2} & \textbf{MAP@3} & \textbf{MAP@4} & \textbf{MAP@5} \\ \midrule 

\netclr  &  0.418  &  0.278  &  0.229 & 0.205  \\
\bapm    &  0.609  &  0.480  &  0.441 & 0.396  \\
\df      &  0.690  &  0.563  &  0.514 & 0.442  \\
\rf      &  0.714  &  0.598  &  0.565 & 0.488  \\
\tiktok  &  0.736  &  0.597  &  0.526 & 0.450  \\
\varcnn  &  0.745  &  0.681  &  0.634 & 0.515  \\
\tmwf   &   0.776  &  0.682  &  0.676 & 0.621 \\
\ours & \textbf{0.927} & \textbf{0.918} & \textbf{0.916} & \textbf{0.914} \\
\bottomrule
\end{tabular}
\end{table}

\subsection{Multi-tab WF Attack in the Open-World}

Now we evaluate the performance of \ours in the open-world scenario. Recall that the open-world experiments regard all non-monitored websites as one website category, while each monitored website is viewed as an individual category. As a result, the number of instances in the non-monitored website category is much larger than that of each monitored category. To avoid the data imbalance problem, we follow the settings in the prior arts~\cite{rimmer2018automated, sirinam2018deep} that mix all closed and open-world instances collected from the same tab setting in our evaluation. For instance, we combine the 2-tab closed and open-world instances to run the 2-tab open-world experiment. 

We measure the \textsf{AUC} scores for monitored and non-monitored websites individually, indicated by $\textsf{AUC}_M$ and $\textsf{AUC}_N$.  
\update{
Figure~\ref{fig:open_auc} demonstrates that \ours achieves the best performance across all tab settings in the open-world scenario. For instance, under the 5-tab setting, \ours achieves the highest $\textsf{AUC}_M$ of 0.9893 and $\textsf{AUC}_N$ of 0.9936, representing average improvements of 25.14\% in $\textsf{AUC}_M$ and 16.95\% in $\textsf{AUC}_N$ over the baselines. Furthermore, Table~\ref{tab:open-mapk} lists the \textsf{MAP@k} results for WF attacks in the open-world scenario, where $k$ represents the number of open tabs. In the 2-tab setting, \ours achieves the best \textsf{MAP@2} of 0.927. Even in the most complex 5-tab setting, \ours maintains a \textsf{MAP@5} above 0.91, with improvements of 345.85\%, 130.81\%, 106.79\%, 87.3\%, 103.11\%, 77.48\%, and 47.18\% over \netclr, \bapm, \df, \rf, \tiktok, \varcnn, and \tmwf, respectively.

As the number of tabs increases, the $\textsf{AUC}_M$ and $\textsf{AUC}_N$ of existing attacks decrease significantly. In contrast, \ours demonstrates robustness in the open-world scenario.
The reason is that \ours utilizes the top-m self-attention mechanism to effectively distinguish monitored websites from non-monitored websites within obfuscated traffic.
}

\noindent \textbf{Remark.} The experimental results in the open-world scenarios demonstrate that \ours is able to accurately identify the monitored websites browsed by the clients even if the complete set of potentially visited websites is unknown a priori. 

\begin{table*}[t]
\centering
\small
\caption{\update{\textsf{AUC}, \textsf{P@k}, and \textsf{MAP@k} of WF attacks on four representative defenses in the 2-tab setting}.}
\label{tab:new_defense}
\begin{tabular}{c|ccc|ccc|ccc|ccc} \toprule

& \multicolumn{3}{c|}{\textbf{Random}}
& \multicolumn{3}{c|}{\textbf{WTF-PAD}} 
& \multicolumn{3}{c|}{\textbf{Front}} 
& \multicolumn{3}{c}{\textbf{RegulaTor}} \\ 
 \cmidrule(lr){2-4} 
 \cmidrule(lr){5-7} 
 \cmidrule(lr){8-10} 
 \cmidrule(lr){11-13}  
\multirow{-2}{*}{} & 
\multicolumn{1}{c}{\textbf{AUC}} & \multicolumn{1}{c}{\textbf{P@2}} & \multicolumn{1}{c|}{\textbf{MAP@2}} & 
\multicolumn{1}{c}{\textbf{AUC}} & \multicolumn{1}{c}{\textbf{P@2}} & \multicolumn{1}{c|}{\textbf{MAP@2}} & 
\multicolumn{1}{c}{\textbf{AUC}} & \multicolumn{1}{c}{\textbf{P@2}} & \multicolumn{1}{c|}{\textbf{MAP@2}} & 
\multicolumn{1}{c}{\textbf{AUC}} & \multicolumn{1}{c}{\textbf{P@2}} & \multicolumn{1}{c}{\textbf{MAP@2}} \\  

\midrule 

\netclr 
& 0.684 & 0.085 & 0.092 
& 0.758 & 0.189 & 0.220 
& 0.662 & 0.061 & 0.064 
& 0.697 & 0.117 & 0.131
\\
\bapm
& 0.893 & 0.384 & 0.464 
& 0.883 & 0.364 & 0.439 
& 0.874 & 0.345 & 0.416
& 0.830 & 0.269 & 0.322
\\
\df
& 0.762 & 0.196 & 0.242 
& 0.889 & 0.436 & 0.544 
& 0.736 & 0.149 & 0.177 
& 0.804 & 0.250 & 0.302
\\
\rf
& 0.601 & 0.100 & 0.115 
& 0.921 & 0.555 & 0.640 
& 0.594 & 0.072 & 0.081 
& 0.863 & 0.367 & 0.436
\\
\tiktok
& 0.779 & 0.221 & 0.268 
& 0.916 & 0.504 & 0.619 
& 0.752 & 0.172 & 0.207 
& 0.849 & 0.315 & 0.382
\\
\varcnn
& 0.796 & 0.273 & 0.329 
& 0.941 & 0.579 & 0.679 
& 0.659 & 0.193 & 0.236 
& 0.834 & 0.273 & 0.325
\\
\tmwf
& 0.962 & 0.657 & 0.730 
& 0.946 & 0.569 & 0.641 
& 0.952 & 0.605 & 0.678 
& 0.904 & 0.409 & 0.477

\\
\ours
& \textbf{0.994} & \textbf{0.888} & \textbf{0.925} 
& \textbf{0.990} & \textbf{0.846} & \textbf{0.893} 
& \textbf{0.991} & \textbf{0.857} & \textbf{0.900} 
& \textbf{0.971} & \textbf{0.709} & \textbf{0.773}

\\
\bottomrule
\end{tabular}
\end{table*}

\subsection{Multi-tab WF Attack under Defenses}

Next, we evaluate the attack performance under WF defenses. Table~\ref{tab:new_defense} lists the experimental results.
\ours achieves robust WF attack performance in the majority of the scenarios and outperforms all baselines. 
\update{
Under the Random defense, \ours achieves the best \textsf{MAP@2} of 0.938, significantly outperforming other WF attacks, all of which have \textsf{MAP@2} values below 0.79. Against the practical WTF-PAD defense, \ours achieves the \textsf{P@2} of 0.846 and the \textsf{MAP@2} of 0.893, representing average improvements of 112.74\% in \textsf{P@2} and 89.73\% in \textsf{MAP@2} over the baselines. 
When facing the SOTA lightweight WF defense, Front, \ours achieves the best \textsf{MAP@2} of 0.9, whereas the \textsf{MAP@2} scores of other WF attacks are less than 0.68.
We observe that all WF attacks have obvious performance degradation against the RegulaTor defense. Note that it is difficult to deploy RegulaTor in the real Tor network due to the data and time overhead.
Even against this defense, \ours still outperforms all baselines with non-trivial margins. For example, \ours achieves the best \textsf{MAP@2} of 0.773, with improvements of 490.08\%, 140.06\%, 155.96\%, 77.29\%, 102.36\%, 137.85\%, and 62.05\% over \netclr, \bapm, \df, \rf, \tiktok, \varcnn, and \tmwf, respectively.

\ours significantly outperforms previous robust WF attacks and multi-tab WF attacks. This superiority is due to the multi-level traffic aggregation of \ours, which effectively extracts robust local features from obfuscated multi-tab traffic under defenses.
}

\noindent \textbf{Remark.} To sum up, these experimental results demonstrate that \ours can perform accurate multi-tab WF attacks even when various WF defenses are present. Compared with the SOTA robust WF attacks, \ours is more resilient against WF defenses. The robustness of \ours is attributed to our \transformer model. It builds robust website fingerprints based on local traffic patterns with less noises and proposes several designs to offset the impacts of noises.

\subsection{Multi-tab Attack under Dynamic Settings}

\begin{table}[t]
\centering
\small
\caption{\update{\textsf{AUC}, \textsf{P@k} and \textsf{MAP@k} of multi-tab WF attacks under dynamic settings}.}

\label{tab:dynamic}
\resizebox{0.48\textwidth}{!}{
\begin{tabular}{c|cccccc}
\toprule
\multirow{2}{*}{} & \multicolumn{3}{c}{\textbf{Multi-tab}} & \multicolumn{3}{c}{\textbf{Defense}} \\ \cmidrule(lr){2-4}  \cmidrule(lr){5-7} 
 & \textbf{AUC} & \textbf{P@5} & \textbf{MAP@5} 
 & \textbf{AUC} & \textbf{P@2} & \textbf{MAP@2} \\ \midrule 
 
\netclr 
& 0.667 & 0.127 & 0.168 
& 0.654 & 0.068 & 0.072
\\
\bapm
& 0.779 & 0.226 & 0.338 
& 0.832 & 0.257 & 0.306
\\
\df
& 0.763 & 0.239 & 0.393 
& 0.769 & 0.199 & 0.238
\\
\rf
& 0.788 & 0.271 & 0.431 
& 0.752 & 0.212 & 0.250
\\
\tiktok
& 0.769 & 0.241 & 0.393 
& 0.793 & 0.228 & 0.279
\\
\varcnn
& 0.809 & 0.289 & 0.467 
& 0.833 & 0.305 & 0.370
\\
\tmwf
& 0.841 & 0.303 & 0.457 
& 0.905 & 0.429 & 0.501
\\
\ours
& \textbf{0.945} & \textbf{0.503} & \textbf{0.707} 
& \textbf{0.987} & \textbf{0.820} & \textbf{0.864}
\\

\bottomrule
\end{tabular}}
\end{table}


\begin{table*}[t]
\centering
\small
\caption{\update{Evaluating the generalization of multi-tab attacks. We compare the \textsf{AUC} scores of multi-tab WF attacks under mismatched numbers of tabs between the training and testing datasets.}}
\label{tab:generalization}
\begin{tabular}{c|ccc|ccc|ccc|ccc} \toprule

& \multicolumn{3}{c|}{\textbf{2-tab (Train)}}
& \multicolumn{3}{c|}{\textbf{3-tab (Train)}} 
& \multicolumn{3}{c|}{\textbf{4-tab (Train)}} 
& \multicolumn{3}{c}{\textbf{5-tab (Train)}} \\ 
 \cmidrule(lr){2-4} 
 \cmidrule(lr){5-7} 
 \cmidrule(lr){8-10} 
 \cmidrule(lr){11-13}  
\multirow{-2}{*}{\textbf{\textbf{\# of tabs (Test)}}} & 
\multicolumn{1}{c}{\textbf{3-tab}} & \multicolumn{1}{c}{\textbf{4-tab}} & \multicolumn{1}{c|}{\textbf{5-tab}} & 
\multicolumn{1}{c}{\textbf{2-tab}} & \multicolumn{1}{c}{\textbf{4-tab}} & \multicolumn{1}{c|}{\textbf{5-tab}} & 
\multicolumn{1}{c}{\textbf{2-tab}} & \multicolumn{1}{c}{\textbf{3-tab}} & \multicolumn{1}{c|}{\textbf{5-tab}} & 
\multicolumn{1}{c}{\textbf{2-tab}} & \multicolumn{1}{c}{\textbf{3-tab}} & \multicolumn{1}{c}{\textbf{4-tab}} \\  

\midrule 

\netclr 
& 0.687 & 0.631 & 0.602 
& 0.750 & 0.629 & 0.596 
& 0.699 & 0.653 & 0.589 
& 0.670 & 0.630 & 0.600
\\
\bapm
& 0.748 & 0.670 & 0.632 
& 0.825 & 0.695 & 0.644 
& 0.745 & 0.728 & 0.640 
& 0.688 & 0.686 & 0.664
\\
\df
& 0.756 & 0.680 & 0.639 
& 0.850 & 0.694 & 0.646 
& 0.808 & 0.741 & 0.642 
& 0.793 & 0.718 & 0.672
\\
\rf
& 0.755 & 0.680 & 0.628 
& 0.850 & 0.705 & 0.639 
& 0.803 & 0.744 & 0.639 
& 0.790 & 0.715 & 0.672
\\
\tiktok
& 0.766 & 0.688 & 0.647 
& 0.863 & 0.707 & 0.657 
& 0.816 & 0.749 & 0.647 
& 0.800 & 0.723 & 0.677
\\
\varcnn
 & 0.794 & 0.717 & 0.664 
 & 0.891 & 0.740 & 0.679 
 & 0.844 & 0.778 & 0.665 
 & 0.835 & 0.762 & 0.713
\\
\tmwf
& 0.758 & 0.672 & 0.631 
& 0.837 & 0.713 & 0.657 
& 0.759 & 0.729 & 0.650 
& 0.729 & 0.689 & 0.667

\\
\ours
& \textbf{0.865} & \textbf{0.777} & \textbf{0.711} 
& \textbf{0.923} & \textbf{0.817} & \textbf{0.740} 
& \textbf{0.891} & \textbf{0.854} & \textbf{0.762} 
& \textbf{0.873} & \textbf{0.817} & \textbf{0.790}

\\
\bottomrule
\end{tabular}
\end{table*}

Next, we evaluate the effectiveness of \ours in the following dynamic settings. (i) Dynamic multi-tab setting: the adversary cannot know the number of tabs opened by the client in advance. (ii) Dynamic defense setting: the adversary has no prior knowledge about the deployed WF defense. 
We use \textsf{AUC}, \textsf{P@k}, and \textsf{MAP@k} to evaluate the performance of multi-tab WF attacks in this section. As listed in Table~\ref{tab:dynamic}, the performance of \ours is superior to all baselines under the dynamic settings. 
\update{
Under dynamic multi-tab settings, \ours achieves an \textsf{AUC} score of 0.945, while other attacks have \textsf{AUC} scores below 0.85. Moreover, \ours demonstrates average improvements of 123.35\% in \textsf{P@5} and 108.56\% in \textsf{MAP@5} compared to the baselines. 
For dynamic defense settings, \ours achieves the best performance with an \textsf{AUC} score of 0.987 and a \textsf{MAP@2} of 0.864. Compared to \netclr, \bapm, \df, \rf, \tiktok, \varcnn, and \tmwf, \ours achieves improvements in \textsf{MAP@2} of 1100\%, 182.35\%, 263.03\%, 245.6\%, 209.68\%, 133.51\%, and 72.46\%, respectively.
We observe that the performance of \netclr is significantly lower than other attacks. The reason is that the data augmentation of \netclr relies on single-tab clean traffic, which is ineffective for multi-tab obfuscated traffic.
}

\noindent \textbf{Remark.} The significant performance improvements above demonstrate the applicability of \ours in practical deployment. This is attributed to our novel architecture of \ours, which can effectively learn the patterns of various websites simultaneously. Thus, \ours achieves more robust website fingerprints regardless of how many tabs are opened or what type of defenses is deployed.

\update{
\subsection{Generalization of Multi-tab Attack}

To evaluate the generalization capability of multi-tab WF attacks, we train models on datasets with 2, 3, 4, and 5 tabs, respectively, and test their performance on datasets with different tab settings.
As shown in Table~\ref{tab:generalization}, we present the AUC scores of multi-tab WF attacks under mismatched training and testing tab settings.
We observe that \ours maintains high \textsf{AUC} scores even under mismatched training and testing tab settings. It significantly outperforms the baselines in such scenarios.
For example, when the number of tabs in the training dataset is 2, 3, 4, or 5, \ours achieves average \textsf{AUC} improvements over the baselines of 14.09\%, 14.2\%, 16.93\%, and 17.07\%, respectively.
While existing attacks experience a performance decline when the number of tabs in the testing dataset exceeds that in the training dataset, \ours shows the smallest degradation. This demonstrates that \ours can effectively generalize across different tab settings.
Based on its performance in current scenarios, \ours is expected to exhibit better generalization on obfuscated traffic with more than 5 tabs compared to baselines.

\noindent \textbf{Remark.} This evaluation highlights the strong generalization capability of \ours. This is mainly due to its multi-level traffic aggregation and top-m self-attention mechanism, which enable \ours to extract stable features under varying tab settings, ensuring excellent performance across diverse scenarios.
}

\begin{figure*}[t]
  \centering
  \includegraphics[width=\linewidth]{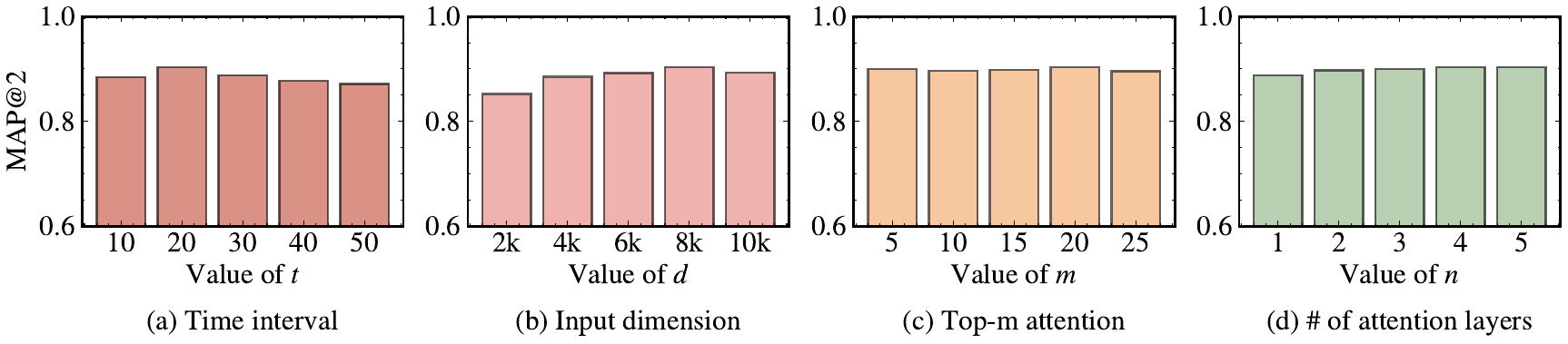}
  \caption{\update{Evaluation of \ours with different parameter settings.}}
  \label{fig:param_analysis}
\end{figure*}
\begin{table}[ht]
\centering
\small
\caption{\update{Ablation analysis of multi-level traffic aggregation module, local analysis module, and website identification module in \ours.}}
\label{tab:ablation-1}
\begin{tabular}{cccc|c}
\toprule
\textbf{Traffic} & \textbf{Local} & \textbf{Website} & & \textbf{MAP@2} \\
\textbf{aggregation} & \textbf{profiling} & \textbf{identification} & & \\
\midrule
\xmark & \cmark & \cmark & & 0.828 \\
\cmark & \xmark & \cmark & & 0.706 \\
\cmark & \cmark & \xmark & & 0.851 \\
\cmark & \cmark & \cmark & & \textbf{0.903} \\
\bottomrule
\end{tabular}
\end{table}

\begin{table}[ht]
\centering
\small
\caption{\update{Ablation analysis of the four aggregation features in the multi-level traffic aggregation module, including packet count, average packet interval, burst count, and average burst size.}}
\label{tab:ablation-2}
\begin{tabular}{ccccc|c}
\toprule
\textbf{Packet} & \textbf{Packet} & \textbf{Burst} & \textbf{Burst} & & \textbf{MAP@2} \\
\textbf{count} & \textbf{interval} & \textbf{count} & \textbf{size} & & \\
\midrule
\cmark & \xmark & \xmark & \xmark & & 0.870 \\
\xmark & \cmark & \xmark & \xmark & & 0.713 \\
\xmark & \xmark & \cmark & \xmark & & 0.838 \\
\xmark & \xmark & \xmark & \cmark & & 0.858 \\
\cmark & \cmark & \cmark & \cmark & & \textbf{0.903} \\
\bottomrule
\end{tabular}
\end{table}

\subsection{Parameter Analysis}
\label{sec:parameter_analysis}

We further study the impact of different parameter values on the performance of \ours. 
\update{
We select four key parameters from the multi-level traffic aggregation module and website identification module, including the aggregation time interval $t$, input dimension $d$, $m$ for top-m attention, and the number of attention layers $n$. 
We evaluate \ours using the 2-tab dataset with the WTF-PAD defense.

As shown in Figure~\ref{fig:param_analysis}, we observe that the performance of \ours is not sensitive to parameter settings. 
When the time interval $t$ is between 10 ms and 50 ms, the \textsf{MAP@2} for \ours varies only between 0.871 and 0.903. Similarly, the differences in \textsf{MAP@2} for different values of $m$ and $n$ are minimal, with only a 0.83\% and 1.78\% variation, respectively. Note that the input dimension $d$ has the most significant impact on \ours. 
When the input dimension $d$ is 2000, \ours achieves a \textsf{MAP@2} of only 0.852, which is due to insufficient website information in the input. As the input dimension $d$ increases, the performance of \ours gradually improves.
}

\noindent \textbf{Remark.} In general, the performance of \ours is not sensitive to parameter choices. The good performance of \ours is attributed to our design rather than carefully crafted parameters.

\subsection{Ablation Analysis}
\label{sec:ablation}

\update{
Next, we perform the ablation analysis of \ours. We evaluate \ours's performance using the 2-tab dataset with the WTF-PAD defense. 
Table~\ref{tab:ablation-1} shows the results of the ablation analysis for the three core modules. We observe that removing any single module leads to a significant drop in \ours's performance. Specifically, when the multi-level traffic aggregation features are replaced with the packet direction sequence, the \textsf{MAP@2} of \ours drops by 8.31\%. 
When the local profiling module is removed, \ours’s \textsf{MAP@2} decreases by 21.82\%, as the top-m self-attention mechanism relies on the local traffic patterns extracted by the local profiling module. Furthermore, replacing the website identification module with an MLP model causes \ours's \textsf{MAP@2} to drop from 0.903 to 0.851.

As shown in Table~\ref{tab:ablation-2}, we analyze the importance of the four aggregation features obtained from the traffic aggregation module. When only packet count, average inter-packet interval, burst count, and average burst size are used as inputs, \ours achieves \textsf{MAP@2} values of 0.87, 0.713, 0.838, and 0.858, respectively. When all four features are used, \ours achieves the optimal \textsf{MAP@2} of 0.903.

\noindent \textbf{Remark.} In summary, all three modules of \ours contribute to its performance. For the multi-level traffic aggregation features, using all four aggregation features provides better robustness compared to using a single aggregation feature.
}

\section{Discussion}
\label{sec:discussion}

\noindent \textbf{Handling Extreme Multi-label Classification.}
Similar to the existing attacks~\cite{sirinam2018deep, sirinam2019triplet, panchenko2016cumul}, our attack considers fingerprinting roughly hundreds of monitored websites. 
If the number of monitored sites increases to tens of thousands or even millions, the multi-tab WF attack problem studied in the paper becomes an extreme multi-label classification (XMLC) problem~\cite{liu2021emerging}.
The existing WF attack methods are unable to solve this problem due to the difficulty of training a holistic model to accurately identify millions of categories. The label tree architecture~\cite{wu2016labeltree, you2018labeltree} may potentially solve this problem since it can hierarchically divide the significantly large label space into smaller subspaces. 
Currently, we do not apply the label tree learning in \ours because its native form cannot well handle the pattern dynamics of different websites. We leave this to our future work.  


\noindent \textbf{Improving Training Performance.}
\ours requires a relatively large time to train classifiers. For example, we need around 60 minutes to train one \transformer with NVIDIA RTX 2080Ti. 
Fortunately, the community has studied the  Transformer training optimization in both natural language processing and computer vision domains~\cite{fang2021turbotransformers, wang2021lightseq}
We can directly apply existing tools, such as Lightseq~\cite{wang2021lightseq} and TurboTransformers~\cite{fang2021turbotransformers}, to accelerate training in \ours. 


\noindent \textbf{Countermeasures against \ours.} The key to reduce the effectiveness of \ours is to reduce the relevance among the website's local patterns. One possible design is as follows. When the Tor exit node loads the page of a website, it first computes the relevance among different HTML elements in the page (\eg by applying Transformer), chooses the elements that are most relevant to others, regularizes them (\eg add dummy data to make them into the same size), and then sends the page's modified traffic to the Tor relay node. Regularizing the most relevant elements blurs the signatures that can be used by \ours. Moreover, it imposes less overhead than regularizing the whole traffic. We leave in-depth exploration of this design to future work.

\section{Related work}
\label{sec:related-work}
\noindent \textbf{Traditional WF Attacks.}
Website fingerprinting (WF) attacks that identify websites visited by clients according to encrypted channels have been extensively studied.
Traditional WF attacks can be classified into two categories: manual feature engineering~\cite{wang2014effective,panchenko2016website,hayes2016k,herrmann2009website,rahman2019tik,oh2019p1} and automated traffic profiling~\cite{abe2016fingerprinting,rimmer2018automated,sirinam2018deep,bhat2019var}. The first category utilizes carefully chosen features and traditional machine learning algorithms. For example, Wang et al.~\cite{wang2014effective} utilized more than 3,000 features to perform the WF attack via the k-Nearest Neighbors (k-NN) classifier.
The CUMUL approach~\cite{panchenko2016website} utilized 104 features and an SVM-based classifier to perform the Internet-scale WF attack. The k-FP attack~\cite{hayes2016k} applied the random forest algorithm to achieve better attack performance than either kNN or CUMUL approach.
The attacks in the second category apply deep learning technologies to construct attacks.  
For instance, Abe and Goto~\cite{abe2016fingerprinting} proposed a Stacked Denoising Autoencoder (SDAE) based WF attack, and Rimmer et al.~\cite{rimmer2018automated} utilized both Convolutional Neural Network (CNN) and Long Short-Term Memory (LSTM) to perform WF attacks. All these existing attacks are unable to accurately fingerprint websites if the monitored traffic includes noises generated by multi-tab browsing or WF defenses.

\noindent \textbf{Robust WF Attacks.} Recently, a series of WF attacks~\cite{sirinam2018deep,rahman2019tik,shen2023RF, mathews2024laserbeak, mitseva2024stop, deng2024ccs} have been proposed to improve the robustness of WF attacks. 
For example, Sirinam et al.~\cite{sirinam2018deep} leveraged sophisticated CNN networks to defeat padding-based defenses, \eg WTF-PAD~\cite{juarez2015wtf}. 
Rahman et al.~\cite{rahman2019tik} proposed Tik-Tok that utilized the features related to the time interval and direction of packets to achieve a robust attack.
\update{
RF~\cite{shen2023RF} and LASERBEAK~\cite{mathews2024laserbeak} further improved feature representation by aggregating traffic features, thereby defeating more advanced defenses.
Mitseva et al.~\cite{mitseva2024stop} enhanced the robustness of WF attacks by jointly analyzing traffic from multiple subpages of the same website. Moreover, Holmes~\cite{deng2024ccs} leveraged the spatiotemporal distribution of traffic to achieve robust attacks during the early stages of webpage loading. 
However, these attacks require a strong assumption that all traffic originates from a single tab, and cannot identify multi-tab obfuscated traffic.
}

\update{
\noindent \textbf{Multi-Tab WF Attacks.} Recently, multi-tab WF attacks have been widely studied~\cite{xu2018multi,yin2021automated,gu2015novel,guan2021bapm, jin2023tmwf, zhao2024oscar} to address the challenges posed by traffic obfuscation when users access multiple websites simultaneously.
Most attacks~\cite{xu2018multi,yin2021automated,gu2015novel} divided the monitored traffic into chunks and fingerprinted the traffic by analyzing the clean chunks without noise.
BAPM~\cite{guan2021bapm} and TMWF~\cite{jin2023tmwf} leverage Transformers to further enhance the performance of multi-tab WF attacks.
Furthermore, Oscar~\cite{zhao2024oscar} employs multi-label metric learning to transform traffic features, effectively enabling fine-grained multi-tab WF attacks.
Unfortunately, they assume that the attacker has prior knowledge about the traffic, \eg the number of tabs (or the maximum number of tabs) in the traffic, which is not practical. Moreover, the performance of existing multi-tab WF attacks significantly degrades under WF defenses.
\ours well addresses these issues, and achieves effective WF attacks in the wild, while being robust to various defenses.
}
\section{Conclusion}
\label{sec:conclusion}

In this paper, we propose \ours, a novel WF attack specifically designed for multi-tab browsing sessions. To fundamentally relax the limitations of existing arts that require prior knowledge on the number of tabs opened in a browsing session, \ours formulates the multi-tab WF attack as a multi-label classification problem. To solve this problem, \ours builds a multi-classifier framework where each classifier is responsible for identifying one specific monitored website. The classifier is designed based on a novel transformer model that can accurately identify websites using local patterns extracted from multiple traffic segments. We implemented a prototype of \ours.
Experimental results demonstrate that \ours significantly improves the performance of multi-tab WF attacks and remains robust even against various WF defenses.

\bibliography{references}

\end{document}